\documentclass[12pt]{article}
\usepackage{graphicx}

\begin{document}

\baselineskip=16pt plus 1pt minus 1pt

\begin{center}

{\large\bf Exactly separable version of X(5) and related models} 

\bigskip\bigskip

{Dennis Bonatsos$^{a}$\footnote{e-mail: bonat@inp.demokritos.gr},
D. Lenis$^{a}$\footnote{e-mail: lenis@inp.demokritos.gr}, 
E. A. McCutchan$^{b}$\footnote{e-mail:
elizabeth.ricard-mccutchan@yale.edu},
D. Petrellis$^{a}$\footnote{e-mail: petrellis@inp.demokritos.gr}, 
I. Yigitoglu$^{c}$\footnote{e-mail: yigitoglu@istanbul.edu.tr} }
\bigskip

{$^{a}$ Institute of Nuclear Physics, N.C.S.R. ``Demokritos'',}

{GR-15310 Aghia Paraskevi, Attiki, Greece}

{$^{b}$ Wright Nuclear Structure Laboratory, Yale University,}

{New Haven, Connecticut 06520-8124, USA}

{$^{c}$ Hasan Ali Yucel Faculty of Education, Istanbul University,}

{TR-34470 Beyazit, Istanbul, Turkey} 

\end{center}

\bigskip\bigskip
\centerline{\bf Abstract} \medskip

One-parameter exactly separable versions of the X(5) and X(5)-$\beta^2$ models,
labelled as ES-X(5) and ES-X(5)-$\beta^2$ respectively, 
are derived by using in the Bohr Hamiltonian potentials of the form 
$u(\beta)+u(\gamma)/\beta^2$. Unlike X(5), in these models the $\beta_1$ and 
$\gamma_1$ bands are treated on equal footing. Spacings within the $\gamma_1$ 
band are well reproduced by both models, while spacings within the $\beta_1$ 
band are well reproduced only by ES-X(5)-$\beta^2$, for which several nuclei 
with $R_{4/2}=E(4_1^+)/E(2_1^+)$ ratios and [normalized to $E(2_1^+$)] 
$\beta_1$ and $\gamma_1$ bandheads corresponding to the model predictions 
have been found.  

\section{Introduction} 

The introduction of the X(5) critical point symmetry \cite{IacX5} has stirred 
considerable effort in studying related special solutions 
\cite{slCaprio,PG,varPLB,BonX5} of the Bohr 
collective Hamiltonian \cite{Bohr}, as well as in identifying nuclei 
exhibiting experimentally \cite{CZX5,Kruecken} 
this behaviour, with considerable success. However, some open questions remain:

1) The separation of variables used in X(5) and related models is approximate. 
In particular, a potential of the form $u(\beta)+u(\gamma)$ is used, 
where $\beta$ and $\gamma$ are the usual collective variables \cite{Bohr}.
In the X(5) model \cite{IacX5} an infinite square well potential is used 
as $u(\beta)$, 
while a harmonic oscillator potential centered around $\gamma =0$ is 
used as $u(\gamma)$. In the X(5)-$\beta^2$ model \cite{BonX5}
a harmonic oscillator potential, $\beta^2/2$, is used as $u(\beta)$. 
Separation of variables is based on two approximations: a) the limitation 
to small angles for $\gamma$, b) the replacement of $\beta^2$ by 
its average value $\langle \beta^2 \rangle$ in the terms involved in the 
$\gamma$-equation. Exact numerical diagonalization of the Bohr Hamiltonian
\cite{Caprio}, carried out using a recently introduced 
computationally tractable version \cite{RoweI,RoweII,RoweIII} 
of the Bohr--Mottelson collective model \cite{Bohr}, pointed out that the 
first approximation is valid for large $\gamma$ stiffness, while the second 
approximation is valid for small $\gamma$ stiffness. 

2) X(5) \cite{IacX5} and the related X(5)-$\beta^2$ model \cite{BonX5}
contain no free 
parameter (up to overall scale factors) in the ground state band and 
$\beta$ bands, but free parameters appear in the $\gamma$ bands and $K=4$
bands. As a result the bandheads of the ground state band 
and the $\beta$ bands, as well as their internal structure, are fixed by the 
theory without any free parameter, while the bandheads of the 
$\gamma$ bands and $K=4$ bands contain free parameters.  
It would have been preferable to treat the $\beta$ and $\gamma$ bands on equal 
footing \cite{IacCam}. 

In the present work we try to circumvent these problems by using potentials 
of the form $u(\beta)+u(\gamma)/\beta^2$, which are known to lead to exact 
separation of variables \cite{Wilets,Fort,Fort2,Heyde}. Then the following 
modifications occur: 

1) The second approximation (replacement of $\beta^2$ by $\langle \beta^2 
\rangle$) is avoided. The first approximation, namely the limitation to small 
angles for $\gamma$, is still used, in order to obtain simplified solutions
of the $\gamma$ equation, but it can be a good one if stiffness is kept
large \cite{Caprio}, which indeed turns out to be the case when comparisons 
to experimental data are performed.

2) The models obtained in this way contain one free parameter in all bands, 
the stiffness of $\gamma$. As a result the relative 
position of all bandheads and the internal structure of all bands 
is fixed by the theory using one parameter, the $\beta_1$ and $\gamma_1$ 
bands treated on equal footing, as it is desirable \cite{IacCam}. 

Recent studies on critical point symmetries \cite{IacX5,slCaprio,PG} 
have made clear that the relative 
position of bandheads in a nucleus, as well as the internal spacing in each 
band, are key structural features which should be reproduced by a model.
The internal spacing of the $\beta_1$ and $\gamma_1$ bands, relative 
to that of the ground state band, will be shown to provide a stringent test 
for the various special solutions of the Bohr Hamiltonian.  

The models occuring from the exact separation of variables will be 
described in Section 2, while in Section 3 some numerical results 
and comparisons to experiment will be given. Finally, a discussion
of the present results and plans for further work will be given in 
Section 4. 

\section{Spectra} 

The original Bohr Hamiltonian \cite{Bohr} is
\begin{equation}\label{eq:e1}
H = -{\hbar^2 \over 2B} \left[ {1\over \beta^4} {\partial \over \partial 
\beta} \beta^4 {\partial \over \partial \beta} + {1\over \beta^2 \sin 
3\gamma} {\partial \over \partial \gamma} \sin 3 \gamma {\partial \over 
\partial \gamma} - {1\over 4 \beta^2} \sum_{k=1,2,3} {Q_k^2 \over \sin^2 
\left(\gamma - {2\over 3} \pi k\right) } \right] +V(\beta,\gamma),
\end{equation}
where $\beta$ and $\gamma$ are the usual collective coordinates, while
$Q_k$ ($k=1$, 2, 3) are the components of angular momentum in the intrinsic 
frame, and $B$ is the mass parameter.  

One seeks \cite{IacX5} solutions of the relevant Schr\"odinger equation having 
the form 
$ \Psi(\beta, \gamma, \theta_i)= \phi_K^L(\beta,\gamma) 
{\cal D}_{M,K}^L(\theta_i)$, 
where $\theta_i$ ($i=1$, 2, 3) are the Euler angles, ${\cal D}(\theta_i)$
denote Wigner functions of them, $L$ are the eigenvalues of angular momentum, 
while $M$ and $K$ are the eigenvalues of the projections of angular 
momentum on the laboratory-fixed $z$-axis and the body-fixed $z'$-axis 
respectively. 

As pointed out in Ref. \cite{IacX5}, in the case in which the potential 
has a minimum around $\gamma =0$ one can write  the angular momentum term 
of Eq. (\ref{eq:e1}) in the form 
\begin{equation}\label{eq:e3} 
\sum _{k=1,2,3} {Q_k^2 \over \sin^2 \left( \gamma -{2\pi \over 3} k\right)}
\approx {4\over 3} (Q_1^2+Q_2^2+Q_3^2) +Q_3^2 \left( {1\over \sin^2\gamma}
-{4\over 3}\right).  
\end{equation}
Using this result in the Schr\"odinger equation corresponding to 
the Hamiltonian of Eq. (\ref{eq:e1}), introducing \cite{IacX5} 
reduced energies 
 $\epsilon = 2B E /\hbar^2$ and reduced potentials $u = 2B V /\hbar^2$,  
and assuming that the reduced potential can be separated into two terms
of the form  $u(\beta, \gamma) = u(\beta) + u(\gamma)/\beta^2$, 
as in Refs. \cite{Wilets,Fort,Fort2,Heyde}, the Schr\"odinger equation can 
be separated into two equations  
\begin{equation} \label{eq:e5}
\left[ -{1\over \beta^4} {\partial \over \partial \beta} \beta^4 
{\partial \over \partial \beta} + {L(L+1)\over 3\beta^2} 
+u(\beta) + {\lambda\over \beta^2} \right] \xi_L(\beta) =
\epsilon  \xi_L(\beta), 
\end{equation}
\begin{equation}\label{eq:e6} 
\left[ -{1\over \sin 3\gamma} {\partial \over 
\partial \gamma}\sin 3\gamma {\partial \over \partial \gamma} 
+{K^2 \over 4}  \left({1\over  \sin^2 \gamma}-{4\over 3}\right)
 +u(\gamma)\right] \eta_K(\gamma) = 
\lambda \eta_K(\gamma).
\end{equation}

Eq. (\ref{eq:e6}) for $\gamma \approx 0$ can be treated as in Ref. 
\cite{IacX5}, considering a potential of the form 
$u(\gamma)= (3c)^2 \gamma^2/2$ and expanding in powers of $\gamma$. 
Then Eq. (\ref{eq:e6}) takes the form 
\begin{equation}
\left[ -{1\over \gamma} {\partial \over \partial \gamma} \gamma
{\partial \over \partial \gamma} + {K^2\over 4\gamma^2} + (3c)^2 
{\gamma^2 \over 2} \right] \eta_K(\gamma) =  \epsilon_\gamma 
\eta_K(\gamma),
\end{equation}
with $\epsilon_\gamma =\lambda +{K^2\over 3}$. 
The solution is given in terms of Laguerre polynomials \cite{IacX5}
\begin{equation}
\epsilon_\gamma = (3c)(n_\gamma+1), \qquad n_\gamma=0,1,2,\ldots,
\end{equation}
\begin{equation}
n_\gamma=0,\quad K=0; \qquad n_\gamma=1, \quad K=\pm 2; \qquad 
n_\gamma=2, \quad K=0,\pm 4; \qquad \ldots, 
\end{equation}
\begin{equation}
\eta_{n_\gamma,K}(\gamma)= C_{n,K} \gamma^{|K/2|} e^{-(3c)\gamma^2/2} 
L_n^{|K/2|} (3c\gamma^2), \qquad n=(n_\gamma -|K/2|)/2.
\end{equation}

Eq. (\ref{eq:e5}) is then solved exactly for the case 
in which $u(\beta)$ is an infinite well potential
\begin{equation}\label{eq:e7} 
 u(\beta) = \left\{ \begin{array}{ll} 0 & \mbox{if $\beta \leq \beta_W$} \\
\infty  & \mbox{for $\beta > \beta_W$} \end{array} \right. .  
\end{equation} 
Setting \cite{IacX5} $\tilde \xi(\beta) = \beta^{3/2} \xi(\beta)$, 
$\epsilon=k_\beta^2$, 
and $z=\beta k_\beta$, one obtains the Bessel equation 
\begin{equation}
{d^2 \tilde \xi \over dz^2 } + {1\over z} {d\tilde \xi \over dz}+
\left[ 1 -{\nu^2 \over z^2}\right] \tilde \xi =0, 
\end{equation}
with 
\begin{equation}
\nu = \sqrt{ {L(L+1)-K^2 \over 3} + {9\over 4} + 3c (n_\gamma+1) }. 
\end{equation}

From the boundary condition $\tilde \xi(\beta_W)=0$ the energy eigenvalues 
are then \cite{IacX5} 
\begin{equation}\label{eq:e8}
\epsilon_{\beta; s,L} = (k_{s,L})^2, \qquad 
k_{s,L}=  {x_{s,L} \over \beta_W},
\end{equation}
where $x_{s,L}$ is the $s$-th zero of the Bessel function 
$J_\nu(k_{s,L}\beta)$, while the relevant eigenfunctions  are
\begin{equation}\label{eq:e10} 
\xi_{s,L}(\beta) = C_{s,L} \beta^{-3/2} J_\nu(k_{s,L} \beta), 
\end{equation}
where $C_{s,L}$ are  normalization constants.
For $K=0$ one has $L=0$, 2, 4, \dots, while for $K\neq 0$ one obtains 
$L=K$, $K+1$, $K+2$, \dots 

The full wave function reads 
\begin{equation}
\Psi(\beta,\gamma,\theta_i) = C_{s,L} \beta^{-3/2} J_\nu(k_{s,L} \beta) 
\eta_{n_\gamma,K}(\gamma) {\cal D}^L_{MK}(\theta_i),
\end{equation}
and should be properly symmetrized \cite{IacX5}
\begin{equation}
\Psi(\beta,\gamma,\theta_i)= {1\over \sqrt{2}} \left[ \phi_{L,K}(\beta,\gamma)
{\cal D}^L _{MK}(\theta_i) + (-1)^{L+K} \phi_{L,-K} 
{\cal D}^L_{M,-K}(\theta_i) \right].  
\end{equation} 

Bands occuring in this model, characterized by ($s$, $n_\gamma$), include 
the ground state band $(1,0)$, the $\beta_1$-band $(2,0)$, the $\gamma_1$-band
$(1,1)$, the first $K=4$ band $(1,2)$. The relative position of all levels 
depends on the single parameter $c$. Therefore the main difference between 
the present model and X(5) is that in the present model all bands are 
fixed by the single parameter $c$, while in X(5) the ground state band 
and the other $n_\gamma=0$ bands are fixed in a parameter-free way, but 
the bandheads of the $n_\gamma \neq 0$ bands depend on free parameters. 

In Ref. \cite{Bijker} a variant of the X(5) model has been considered, 
in which during the separation of variables the term $K^2/3$ has been 
kept in the $\beta$-equation, while in Ref. \cite{IacX5} this term has 
been put in the $\gamma$-equation. This choice leads to different results 
(different expression for $\nu$, in particular) when the method of 
Refs. \cite{IacX5,Bijker} is followed, but it makes no difference in the 
present approach. 

Eq. (\ref{eq:e5}) is exactly soluble also in the case in which 
$u(\beta)= \beta^2/2$. In this case, which is analogous to 
the X(5)-$\beta^2$ model \cite{BonX5}, the eigenfunctions are \cite{Mosh1555}
\begin{equation}\label{eq:e11}
F_n^L(\beta)= \left[ {2 n!\over \Gamma 
\left(n+a+{5\over 2}\right)}\right]^{1/2} \beta^a L_n^{a+{3\over 2}}(\beta^2)
e^{-\beta^2/2}, 
\end{equation} 
where $\Gamma(n)$ stands for the $\Gamma$-function, $L_n^a(z)$ denotes the 
Laguerre polynomials, and 
\begin{equation}\label{eq:e12} 
a= -{3\over 2}+\sqrt{ {L(L+1)-K^2\over 3}+{9\over 4} + 3c(n_\gamma+1) }, 
\end{equation} 
while the energy eigenvalues are 
\begin{equation}\label{eq:e13} 
E_{n,L}= 2n+a+{5\over 2}= 2n+1 + \sqrt{ { L(L+1)-K^2\over 3} +{9\over 4} +
3c (n_\gamma+1) }, 
\qquad n=0,1,2,\ldots 
\end{equation}

In the above, $n$ is the usual oscillator quantum number. 
A formal correspondence between the energy levels of the 
X(5) analogue and the present X(5)-$\beta^2$ analogue 
can be established through the relation 
\begin{equation}\label{eq:e14}
n=s-1.
\end{equation} 
It should be remembered, however, that 
the origin of the two quantum numbers is different, $s$ labelling the 
order of a zero of a Bessel function and $n$ labelling the number of zeros 
of a Laguerre polynomial. In the present notation, the ground state band 
corresponds to $s=1$ ($n=0$). For the energy states the notation 
$E_{s,L} = E_{n+1,L}$ of Ref. \cite{IacX5} will be kept.  

On the present approach the following general comments apply. 

a) The use of a potential of the form $u(\beta)+u(\gamma)/\beta^2$, instead 
of a potential of the form $u(\beta)+u(\gamma)$ [as in X(5) 
and X(5)-$\beta^2$] leads to exact separation of variables instead of an 
approximate one \cite{Wilets,Fort,Fort2,Heyde}. 
As a result, no $\beta^2$ factors appear in the 
$\gamma$-equation, and therefore the approximation of replacing $\beta^2$ 
by its average value, $\langle \beta^2 \rangle$, used in X(5) and 
X(5)-$\beta^2$, is avoided. Exact numerical diagonalizations 
\cite{Caprio} of the Bohr Hamiltonian have demonstrated that this 
approximation is valid only for small $\gamma$ stiffness. This requirement 
is removed in the present case.    

b) However, the treatment of the $\gamma$-equation in the present approach 
is based on the same approximation of small $\gamma$ angles also used in the 
X(5) and X(5)-$\beta^2$ models. The potential $(3c)^2 \gamma^2 /2$ used here 
is the lowest order approximation for small $\gamma$ to the potential 
$c^2 (1-\cos 3\gamma)$,  which has also been used in Ref. \cite{Jean}. 
It should be reminded that the dependence on $\cos 3\gamma$ results from 
the symmetry requirements \cite{Bohr} of the Bohr Hamiltonian, explicitly 
listed in Ref. \cite{Corrigan}. A two-dimensional oscillator in $\gamma$,
similar to the one obtained here, has also been obtained in Ref. \cite{Jean}
in the limit of large $\gamma$ stiffness (large $c$ in the present notation). 
The exact numerical diagonalizations of the Bohr Hamiltonian carried out 
in Ref. \cite{Caprio} consistently  demonstrated that the small angle 
approximation for $\gamma$ is good for large $\gamma$ stiffness, which 
in the present models can be achieved, since the requirement of small
$\gamma$ stiffness is not present any more, as discussed in point a).   
In Sec. 3 we shall see that experimental data are reproduced for 
values of $c$ of order 10, corresponding to $(3c)^2 \approx 900$.  

c) Small oscillations
in $\gamma$ around the zero value, corresponding to axially deformed prolate 
shapes, have also been considered in Ref. \cite{Davydov}, leading to the 
conclusion that $K$ can be considered as a good quantum number either if 
$\gamma$ is fixed to zero, or if the nucleus is strongly deformed.   
In Sec. 3 we shall see that good agreement with experimental data 
is obtained for nuclei with $R_{4/2}=E(4_1^+)/E(2_1^+) > 3.15$, 
i.e. for nuclei which are well deformed.  

\section{Numerical results and comparison to experiment} 

Numerical results for the present models, referred to as {\sl exactly 
separable X(5)} [ES-X(5)] and {\sl exactly separable X(5)-$\beta^2$} 
[ES-X(5)-$\beta^2$] respectively, are shown in Table 1, together with results 
for several other models, including X(5) \cite{IacX5}, exact numerical 
diagonalization of the Bohr Hamiltonian \cite{Caprio} (labelled by 
``Caprio''), 
X(5)-$\beta^{2n}$ ($n=1,2,3,4$) \cite{BonX5}, 
X(5) with a Davidson potential 
\begin{equation}\label{eq:e20}
u(\beta)=\beta^2 +{\beta_0^4 \over \beta^2}, 
\end{equation}
where $\beta_0$ is the minimum of the potential \cite{varPLB}, labelled 
as X(5)-D. The collective 
quantities reported in Table 1 include the ground state band 
ratio $R_{4/2}=E(4_1^+)/E(2_1^+)$, 
the bandheads of the $\beta_1$ and $\gamma_1$ bands, $E(0_\beta^+)$ 
and $E(2_\gamma^+)$,  normalized to $E(2_1^+)$, the spacings within 
the $\beta_1$ band relative to these of the ground state band
\begin{equation}\label{eq:e21}
R_{2,0,\beta,g}= {E(2_\beta^+)-E(0_\beta^+) \over E(2_1^+)},\qquad    
R_{4,2,\beta,g}= {E(4_\beta^+)-E(2_\beta^+) \over E(4_1^+)-E(2_1^+)}, 
\end{equation}
and the spacing within the $\gamma_1$ band relative to that of the 
ground state band 
\begin{equation}\label{eq:e22} 
R_{4,2,\gamma,g}={E(4_\gamma^+)-E(2_\gamma^+) \over E(4_1^+)-E(2_1^+)}.
\end{equation}
Experimental values for the energy ratios $R_{2,0,\beta,g}$, 
$R_{4,2,\beta,g}$, and $R_{4,2,\gamma,g}$ are shown in Fig. 1 for all nuclei
with $A > 50$ (excluding magic and semimagic nuclei). 
The following comments can be made:

1) From Fig. 1(a),(b) it is clear that the majority of nuclei exhibit 
ratios $R_{2,0,\beta,g}$ and $R_{4,2,\beta,g}$ close to 1 or slightly below 
it, indicating that the spacings within the $\beta_1$ band are similar 
to the spacings within the ground state band, as expected for a band 
displaced from the ground state band by one quantum of $\beta$ vibration.  
Ratios exactly equal to 1 are provided by the X(5)-$\beta^2$, 
ES-X(5)-$\beta^2$, and X(5)-D models, guaranteed by the $\beta^2$ term 
present in the relevant potentials. X(5) gives values of 1.8 and 1.7
respectively, the well known point of disagreement with experimental ratios 
by a factor close to 2 \cite{CZX5,Kruecken}. The exact numerical 
diagonalization 
of Ref. \cite{Caprio} provides similar or higher values, while ES-X(5)
gives values around 1.5~. The X(5)-$\beta^4$, X(5)-$\beta^6$, and 
X(5)-$\beta^8$ models interpolate between X(5)-$\beta^2$ and X(5), 
as expected, since higher powers of $\beta^{2n}$ closer approximate the 
infinite well potential.

2) The above observations can be understood in the following way. 
It is known that the problem of overprediction of the spacing 
within the $\beta_1$ band by X(5) can be resolved by replacing the infinite 
well potential in $\beta$ by a potential with sloped walls \cite{slCaprio}.
The combination of the five-dimensional centrifugal term with the 
sloped well provides a potential with a minimum, resembling the Davidson 
potential of Eq. (\ref{eq:e20}) as well as the sum of a harmonic oscillator 
potential and a centrifugal term.  

3) From Fig. 1(c) it is clear that the majority of nuclei exhibit 
ratios $R_{4,2,\gamma,g}$ close to 1, indicating that the spacings within 
the $\gamma_1$ band are similar to the spacings within the ground state 
band. Among the models of Table 1, X(5), X(5)-$\beta^{2n}$, as well as X(5)-D, 
provide values slightly higher than 1, the ES-X(5) and ES-X(5)-$\beta^2$ 
models give values slightly lower than 1, while
the exact numerical diagonalization of Ref. \cite{Caprio} is in between. 
It is interesting that in the X(5)-D model for large 
parameter values, the spacing within the $\gamma_1$ band becomes the same as 
within the ground state band and the $\beta_1$ bands, as expected in the 
SU(3) limit. 

From the above observations it is expected that the one-parameter 
ES-X(5)-$\beta^2$ and X(5)-D models, as well as X(5)-$\beta^2$, are more 
appropriate for reproducing the correct spacings within the $\beta_1$ 
and $\gamma_1$ bands. However, the position of the bandheads is also 
important. It is then reasonable to look for nuclei for which a model 
can closely reproduce the $R_{4/2}$ ratio, characterizing the development of
the ground state band, but also the development of the $\beta_1$ and 
$\gamma_1$ bands, according to the systematics of Fig. 1, as well 
as the normalized bandheads $E(0_\beta^+)/E(2_1^+)$ and 
$E(2_\gamma^+)/E(2_1^+)$. A search of all even nuclei with $Z>50$, for which 
sufficient data exist \cite{NDS}, provided the results shown in Table 2. 
18 examples have been found for ES-X(5)-$\beta^2$, as well as 4 examples 
for ES-X(5). 

The basic difference between ES-X(5)-$\beta^2$ and ES-X(5) is shown 
in Table 3, where the ground state, $\beta_1$ and $\gamma_1$ bands 
of $^{156}$Gd, a good example of ES-X(5)-$\beta^2$, and $^{162}$Dy, 
a good example of ES-X(5), are shown. In the first case the agreement 
between theory and experiment remains good in all three bands up to high 
angular momenta, while in the second case this holds only for the ground 
state and $\gamma_1$ bands, the theoretical $\beta_1$ band diverging 
from the data with increasing angular momentum. 

A final remark concerning the physical content of the models developed 
here, based on the potential of the form $u(\beta)+u(\gamma)/\beta^2$
\cite{Wilets,Fort,Fort2,Heyde}. As already remarked in Ref. \cite{Wilets},
one expects $\gamma$ stability to increase with deformation. This is 
corroborated by the results shown in Table 2, where it is clear that $c$
is increasing with $R_{4/2}$, indicating that $u(\gamma)$ is becoming 
more and more steep with increasing deformation, restricting the nucleus 
to $\gamma$ values very close to zero. As a result, the present models 
are applicable to strongly deformed nuclei close to axial symmetry. 

\section{Discussion}

In summary, exactly separable one-parameter versions of the X(5) and 
X(5)-$\beta^2$ models, labelled as ES-X(5) and ES-X(5)-$\beta^2$,
have been derived, by using potentials of the 
form $u(\beta)+u(\gamma)/\beta^2$. Unlike X(5), in these models the $\beta_1$ 
and $\gamma_1$ bands are treated on equal footing. The spacings within 
the $\gamma_1$ band are in agreement to experimental evidence in both models,
while the spacings within the $\beta_1$ band are reproduced correctly only 
by ES-X(5)-$\beta^2$. Several nuclei for which the $R_{4/2}$ ratio, as well
as the normalized positions of the $\beta_1$ and $\gamma_1$ bandheads are 
closely reproduced by ES-X(5)-$\beta^2$ have been identified. A detailed study
of the complete level schemes of these nuclei, including B(E2) transition 
rates, is deferred to a longer publication.

\newpage  

\begin{table}

\caption{The key collective quantities $R_{4/2}=E(4_1^+)/E(2_1^+)$, 
normalized $\beta_1$ bandhead $E(0_\beta^+)/E(2_1^+)$ (labelled as
$0_\beta^+/2_1^+$), normalized $\gamma_1$ bandhead $E(2_\gamma^+)/E(2_1^+)$
(labelled as $2_\gamma^+/2_1^+$), spacings of the $\beta_1$ band relative 
to the ground state band $R_{2,0,\beta,g}$ [Eq. (\ref{eq:e21})] and 
$R_{4,2,\beta,g}$ [Eq. (\ref{eq:e21})], as well as spacing of the $\gamma_1$
band relative to the ground state band $R_{4,2\gamma,g}$ [Eq. (\ref{eq:e22})]
are listed for different models, including X(5)-$\beta^{2n}$ \cite{BonX5}, 
X(5) \cite{IacX5}, X(5)-D \cite{varPLB}, Caprio's exact numerical 
diagonalization of the Bohr Hamiltonian \cite{Caprio}, and the present exactly 
separable analogues of X(5) [ES-X(5)] and X(5)-$\beta^2$ [ES-X(5)-$\beta^2$].
Parameter values, if present, are listed in the column ``Par'',
using the definition and symbol of the relevant original publication. 
The notation ``par'' indicates that the corresponding quantity 
depends on an additional free parameter.   
}

\bigskip

\begin{tabular}{ l r r r r  r r r }
\hline
model & Par & $R_{4/2}$ & $0_{\beta}^+/2_1^+$ & $2_{\gamma}^+/2_1^+$ &
$R_{2,0,\beta,g}$ & $R_{4,2,\beta,g}$ & $R_{4,2,\gamma,g}$ \cr 
\hline
   &   &     &       &   &  &  &       \\
X(5)-$\beta^2$ & & 2.646 & 3.562 & par & 1.000 & 1.000 & 1.132 \\ 
X(5)-$\beta^4$ & & 2.769 & 4.352 & par & 1.250 & 1.205 & 1.101 \\
X(5)-$\beta^6$ & & 2.824 & 4.816 & par & 1.416 & 1.344 & 1.089 \\
X(5)-$\beta^8$ & & 2.852 & 5.091 & par & 1.528 & 1.441 & 1.083 \\
X(5)           & & 2.904 & 5.649 & par & 1.801 & 1.701 & 1.071 \\
\hline
X(5)-D & $\beta_0$ & & & & & & \\
 & 0.0 & 2.646 & 3.562 & par & 1.000 & 1.000 & 1.131 \\
 & 1.0 & 2.756 & 4.094 & par & 1.000 & 1.000 & 1.108 \\
 & 1.5 & 2.978 & 5.756 & par & 1.000 & 1.000 & 1.064 \\
 & 2.0 & 3.156 & 8.772 & par & 1.000 & 1.000 & 1.031 \\
 & 5.0 & 3.327 &50.130 & par & 1.000 & 1.000 & 1.001 \\ 
\hline 
Caprio & a & & & & & & \\
 & 0.   & 2.20 & 3.03 & 2.20 & 1.77 & 0.31 & 1.16 \\
 & 200. & 2.76 & 5.66 & 6.09 & 2.31 & 1.74 & 1.15 \\
 & 400. & 3.02 & 8.37 &10.12 & 2.19 & 1.76 & 1.04 \\
 & 600. & 3.12 &10.26 &13.42 & 2.00 & 1.79 & 0.95 \\
 & 800. & 3.17 &11.71 &16.19 & 1.91 & 1.70 & 1.00 \\
 &1000. & 3.20 &12.89 &18.66 & 1.84 & 1.69 & 0.97 \\
\hline 
ES-X(5) & c & & & & & & \\
 & 2.0 & 3.166 & 10.298 &  3.166 & 1.649 & 1.606 & 0.929 \\
 & 4.0 & 3.234 & 13.643 &  5.955 & 1.579 & 1.552 & 0.909 \\
 & 6.0 & 3.264 & 16.451 &  8.764 & 1.534 & 1.515 & 0.904 \\
 & 8.3 & 3.283 & 19.292 & 12.013 & 1.497 & 1.484 & 0.903 \\
 &10.0 & 3.292 & 21.210 & 14.423 & 1.477 & 1.465 & 0.904 \\
 &12.0 & 3.299 & 23.317 & 17.266 & 1.456 & 1.447 & 0.905 \\
 &13.7 & 3.304 & 25.012 & 19.692 & 1.442 & 1.433 & 0.906 \\
\hline
ES-X(5)-$\beta^2$ & c & & & & & & \\
 & 2.0 & 3.006 &  6.074 &  3.006 & 1.000 & 1.000 & 0.852 \\
 & 4.0 & 3.117 &  7.806 &  5.516 & 1.000 & 1.000 & 0.796 \\
 & 6.0 & 3.171 &  9.217 &  8.011 & 1.000 & 1.000 & 0.771 \\
 & 8.0 & 3.204 & 10.439 & 10.502 & 1.000 & 1.000 & 0.757 \\
 &10.0 & 3.225 & 11.531 & 12.991 & 1.000 & 1.000 & 0.749 \\
 &12.0 & 3.241 & 12.529 & 15.478 & 1.000 & 1.000 & 0.742 \\
 &14.0 & 3.252 & 13.453 & 17.965 & 1.000 & 1.000 & 0.738 \\
\hline
\end{tabular}
\end{table}

\newpage 
\begin{table}

\caption{Comparison of theoretical predictions of the exactly separable
analogue of the X(5)-$\beta^2$ model [ES-X(5)-$\beta^2$] (upper part) 
and of the exactly separable analogue of the X(5) model [ES-X(5)]
(lower part) to experimental $R_{4/2}=E(4_1^+)/E(2_1^+)$
ratios, as well as to experimental $\beta_1$ and $\gamma_1$ bandheads, 
normalized to the $2_1^+$ state and labelled by $0_\beta^+/2_1^+$ and 
$2_\gamma^+/2_1^+$ respectively. All data have been taken from Ref. 
\cite{NDS}. 
}

\bigskip

\begin{tabular}{ r r r r | r r r r}
\hline
nucleus & $R_{4/2}$ & $0_{\beta}^+/2_1^+$ & $2_{\gamma}^+/2_1^+$ &
$c$     & $R_{4/2}$ & $0_{\beta}^+/2_1^+$ & $2_{\gamma}^+/2_1^+$ \\
        & exp &  exp  & exp &  & th & th & th \\
\hline
   &       &       &   &  &  &       &\\
$^{188}$Os & 3.083 &  7.008 &  4.083 &  2.9 & 3.068 &  6.908 &  4.139 \\
$^{186}$Os & 3.165 &  7.736 &  5.596 &  4.0 & 3.117 &  7.806 &  5.516 \\
$^{184}$Os & 3.203 &  8.698 &  7.870 &  5.7 & 3.165 &  9.019 &  7.637 \\
$^{184}$W  & 3.274 &  9.014 &  8.122 &  6.0 & 3.171 &  9.217 &  8.011 \\
$^{162}$Er & 3.230 & 10.654 &  8.827 &  7.0 & 3.189 &  9.847 &  9.257 \\
$^{166}$Yb & 3.228 & 10.189 &  9.108 &  7.0 & 3.189 &  9.847 &  9.257 \\ 
$^{158}$Dy & 3.206 & 10.014 &  9.567 &  7.3 & 3.194 & 10.028 &  9.630 \\
$^{170}$Er & 3.310 & 11.335 & 11.883 &  9.2 & 3.218 & 11.107 & 11.995 \\
$^{182}$W  & 3.291 & 11.346 & 12.201 &  9.4 & 3.220 & 11.215 & 12.244 \\
$^{180}$Hf & 3.307 & 11.807 & 12.855 & 10.0 & 3.225 & 11.531 & 12.991 \\
$^{156}$Gd & 3.239 & 11.796 & 12.972 & 10.0 & 3.225 & 11.531 & 12.991 \\
$^{228}$Ra & 3.207 & 11.300 & 13.258 & 10.1 & 3.226 & 11.583 & 13.115 \\
$^{170}$Yb & 3.293 & 12.692 & 13.598 & 10.7 & 3.231 & 11.890 & 13.861 \\
$^{230}$Th & 3.273 & 11.934 & 14.687 & 11.3 & 3.236 & 12.189 & 14.608 \\
$^{228}$Th & 3.235 & 14.402 & 16.776 & 13.4 & 3.249 & 13.182 & 17.219 \\
$^{154}$Sm & 3.254 & 13.410 & 17.567 & 13.7 & 3.251 & 13.318 & 17.592 \\
$^{172}$Yb & 3.305 & 13.245 & 18.616 & 14.4 & 3.254 & 13.630 & 18.463 \\
$^{232}$U  & 3.291 & 14.530 & 18.221 & 14.5 & 3.255 & 13.674 & 18.587 \\
\hline
   &       &       &   &    &  &   &\\
$^{166}$Er & 3.289 & 18.118 &  9.754 &  6.7 & 3.271 & 17.352 &  9.751 \\
$^{162}$Dy & 3.294 & 17.332 & 11.011 &  7.6 & 3.278 & 18.462 & 11.022 \\
$^{162}$Gd & 3.291 & 19.792 & 11.983 &  8.3 & 3.283 & 19.292 & 12.013 \\
$^{176}$Yb & 3.308 & 21.661 & 15.352 & 10.6 & 3.294 & 21.857 & 15.275 \\
\hline
\end{tabular}
\end{table}

 \newpage 
\begin{table}

\caption{Comparison of the predictions of the exactly separable analogue 
of the X(5)-$\beta^2$ model [ES-X(5)-$\beta^2$] (with $c=10.0$)
to the experimental spectrum of $^{156}$Gd \cite{NDS} (upper part), 
and of the predictions of the exactly separable analogue of X(5) [ES-X(5)] 
(with $c=7.6$) to the experimental spectrum of $^{162}$Dy \cite{NDS} 
(lower part). The energy levels of the ground state band (gsb), the $\beta_1$ 
band, and the $\gamma_1$ band are normalized to the $2_1^+$ level.
}

\bigskip

\begin{tabular}{ r r r r r r | r r r }
\hline
  &   & gsb    &  gsb &$\beta_1$&$\beta_1$& &$\gamma_1$&$\gamma_1$\\
  &L  &  exp   &  th    &  exp   &  th    &L  & exp    & th     \\
$^{156}$Gd & & &        &        &        &   &        &        \\
  &0  &  0.000 &  0.000 & 11.796 & 11.531 & 2 & 12.972 & 12.991 \\
  &2  &  1.000 &  1.000 & 12.695 & 12.531 & 3 & 14.027 & 13.712 \\
  &4  &  3.239 &  3.225 & 14.587 & 14.757 & 4 & 15.235 & 14.656 \\
  &6  &  6.573 &  6.468 & 17.311 & 17.999 & 5 & 16.937 & 15.811 \\
  &8  & 10.848 & 10.500 & 20.775 & 22.031 & 6 & 18.474 & 17.162 \\
  &10 & 15.916 & 15.122 & 24.952 & 26.653 & 7 & 20.792 & 18.692 \\
  &12 & 21.631 & 20.179 & 30.435 & 31.710 & 8 & 22.607 & 20.388 \\
  &14 & 27.828 & 25.559 &        &        & 9 & 25.285 & 22.233 \\
  &16 & 34.388 & 31.180 &        &        &10 & 27.452 & 24.213 \\
$^{162}$Dy & & &        &        &        &   &        &        \\
  &0  &  0.000 &  0.000 & 17.332 & 18.462 & 2 & 11.011 & 11.022 \\
  &2  &  1.000 &  1.000 & 18.020 & 19.969 & 3 & 11.938 & 11.909 \\
  &4  &  3.294 &  3.278 & 19.518 & 23.369 & 4 & 13.154 & 13.081 \\
  &6  &  6.800 &  6.733 & 21.913 & 28.445 & 5 & 14.664 & 14.532 \\
  &8  & 11.412 & 11.259 & 24.619 & 34.975 & 6 & 16.420 & 16.255 \\
  &10 & 17.044 & 16.768 & 28.041 & 42.778 & 7 & 18.477 & 18.242 \\
  &12 & 23.572 & 23.195 & 32.156 & 51.721 & 8 & 20.709 & 20.485 \\
  &14 & 30.895 & 30.493 & 36.640 & 61.705 & 9 & 23.284 & 22.975 \\
  &16 & 38.904 & 38.625 &        &        &10 & 25.879 & 25.708 \\
  &18 & 47.583 & 47.568 &        &        &11 & 28.981 & 28.676 \\
  &   &        &        &        &        &12 & 31.396 & 31.873 \\
  &   &        &        &        &        &13 & 35.453 & 35.295 \\
  &   &        &        &        &        &14 & 39.437 & 38.936 \\
\hline
\end{tabular}
\end{table}

\newpage 

{\bf Figures}

\begin{figure}[htb]
\includegraphics[width=50mm]{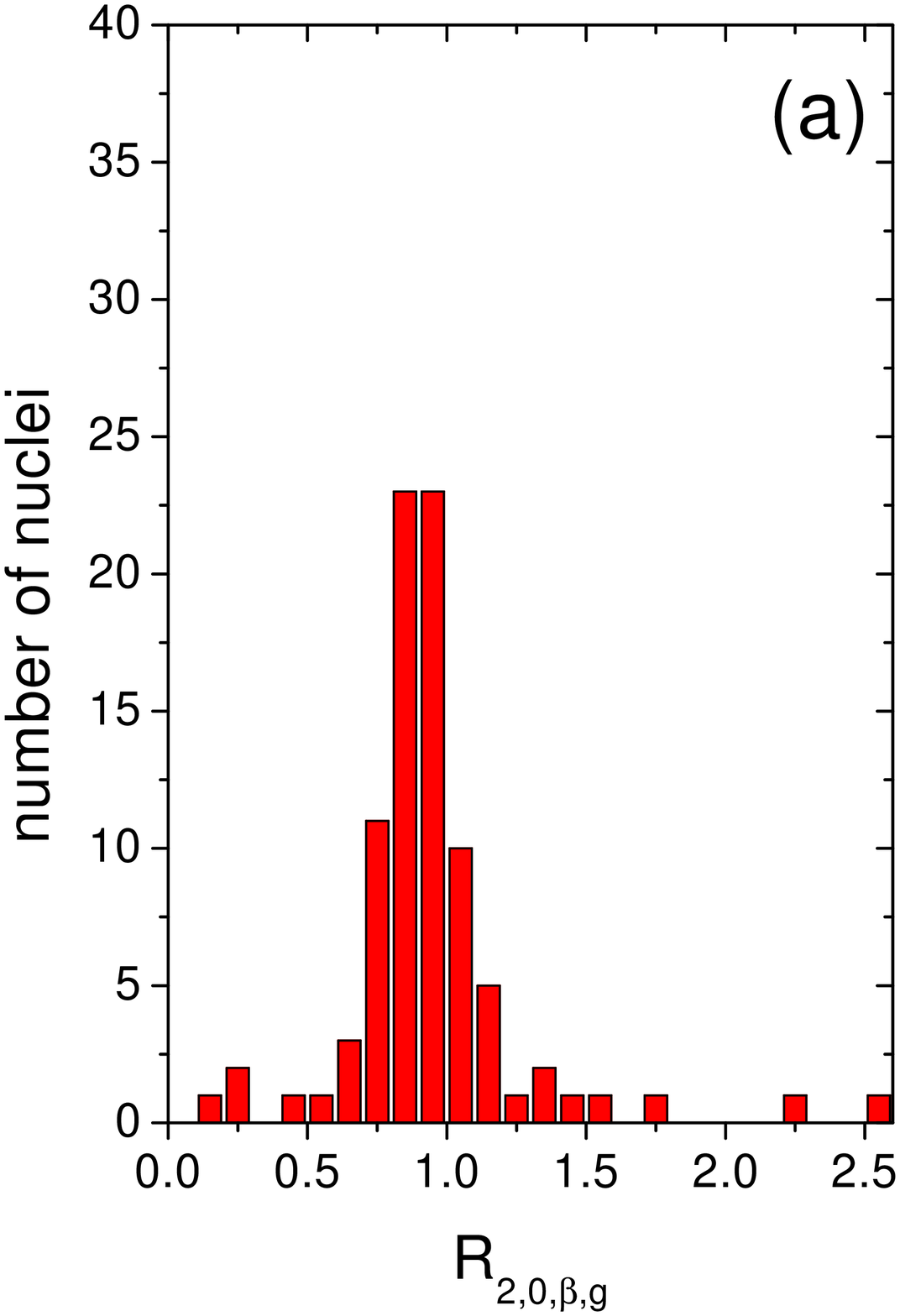}\hspace{5mm}
\includegraphics[width=50mm]{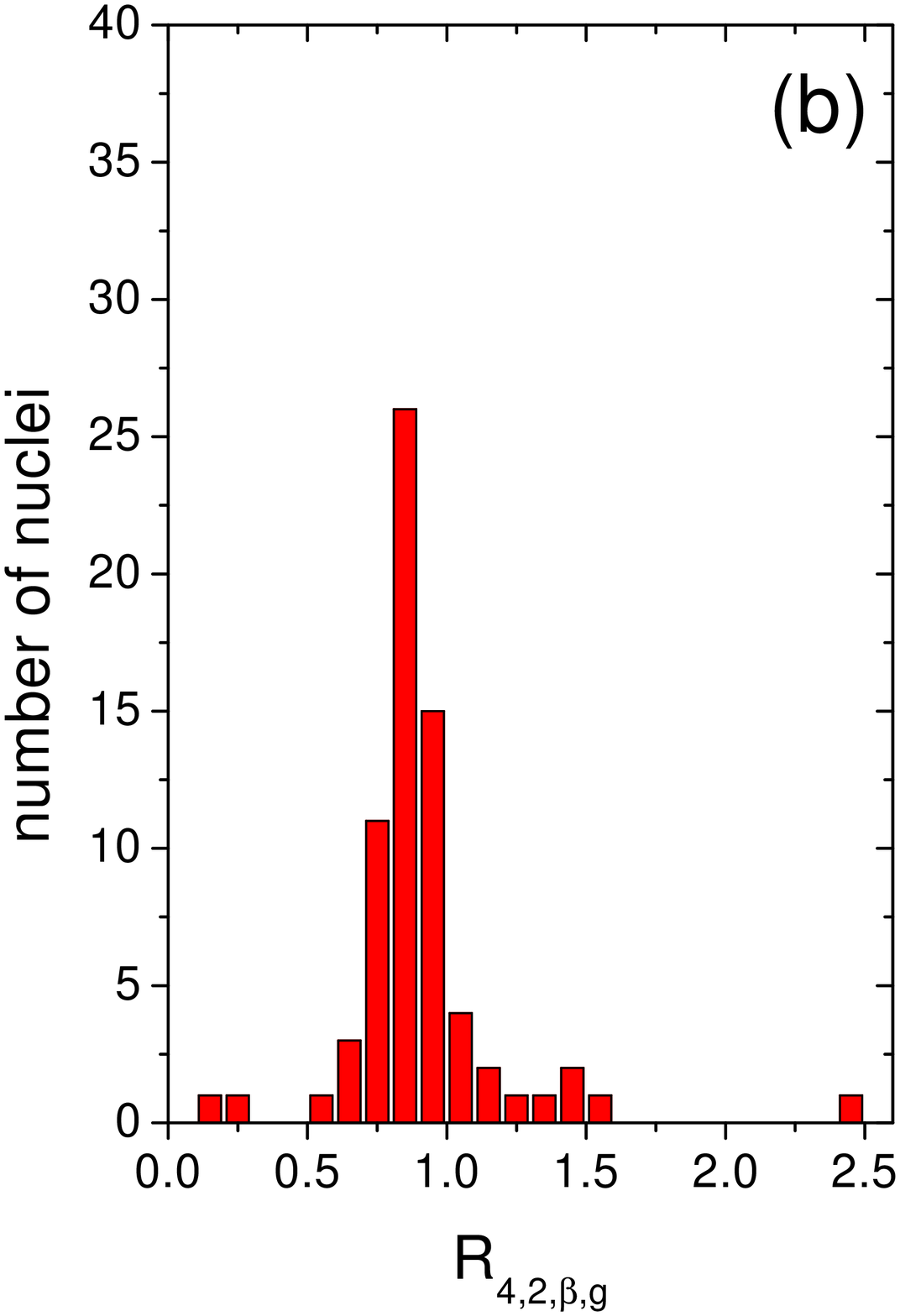}\hspace{5mm}
\includegraphics[width=50mm]{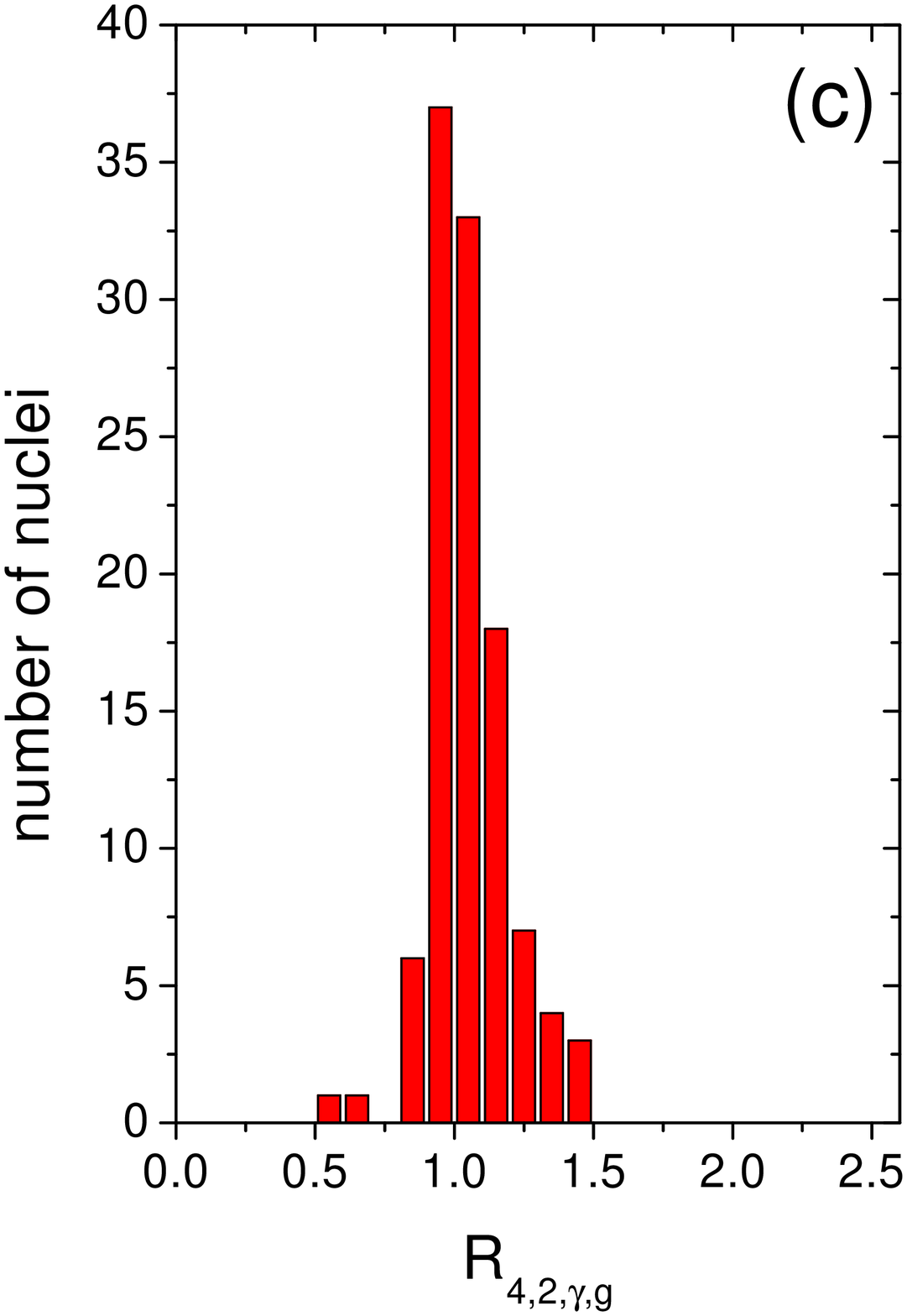}
\caption{ Experimental data for the energy ratios 
$R_{2,0,\beta,g}$  [Eq. (\ref{eq:e21})] (a), 
$R_{4,2,\beta,g}$  [Eq. (\ref{eq:e21})] (b), and
$R_{4,2,\gamma,g}$ [Eq. (\ref{eq:e22})] (c).
For each ratio, all nuclei with 
$A>50$ (excluding magic and semimagic nuclei) for which sufficient 
experimental data (taken from Ref. \cite{NDS}) exist, have been taken into 
account.     } 
\end{figure}

\end{document}